\documentstyle[multicol,aps,epsf]{revtex}
\begin{document}
\title{Disorder-Driven Magnetic Field Dependence of the
Internal Field Distribution in the Bragg Glass Phase 
of Type-II Superconductors}
\author{Mohammad Kohandel$^1$, and 
Michel J. P. Gingras$^{1,2}$}
\address{$^1$ Department of Physics, University of Waterloo, 
Waterloo, Ontario, N2L 3G1, Canada \\
$^2$ Canadian Institute for Advanced Research, 
180 Dundas Street West, Toronto, Ontario, M5G 1Z8, Canada}
\date{\today}
\maketitle
\begin{abstract}
We use the replica variational method to study 
the effects of weak point disorder on the 
variance of the internal field distribution measured 
in NMR and muon-spin rotation experiments in 
type-II superconductors.
We show that for a simple model there is significant 
magnetic field dependence which is extrinsic and 
disorder-driven, and does not have a microscopic 
(non $s-$wave pairing) origin. Results are presented 
where we examine the dependence of the magnetic field 
variance upon the strength of the applied external field. 

\vspace{0.2cm}
\end{abstract}
\begin{multicols}{2}

The study of high temperature superconductors (high-$T_c$'s) 
remains a subject of intense
investigation, both in the search of an explanation 
of its microscopic origin and in the application of
phenomenological theories 
to investigate the effects of thermal fluctuations, random
disorder, and transport.
It is now generally accepted that high-$T_c$
superconductors 
are unconventional $d-$wave rather than conventional 
$s-$wave with an energy gap that vanishes along certain
directions in momentum space, resulting in 
nodes in the superconducting gap. Measurement of 
the temperature dependence of the magnetic
penetration depth, $\lambda(T)$, is one way to
probe the non $s-$wave nature of the low energy nodal excitations
of the superconducting state \cite{Hardy}.
A detailed understanding of the $d-$wave nature
of the high-$T_c$'s  
must also include a quantitative description
of the dependence of the two fundamental length scales   
characterizing a superconductor, 
the penetration depth $\lambda$ and the 
coherence length $\xi_0$, on the strength of an
applied magnetic field $B_0$.
An early theoretical investigation of the weak-field
response of a d$_{x^2-y^2}$ superconductor predicted
a direction-dependent nonlinear Meissner effect,
associated with the quasiclassical shift of the 
excitation spectrum due to the superflow created
by the screening currents \cite{Yip}. 
More recently, Amin {\it et al.} 
studied nonlinear and nonlocal effects due to
the field induced excitations at the gap nodes to predict the
temperature and field dependence of an effective
penetration depth $\lambda^{\rm eff}$ \cite{Amin}. 

While understanding the microscopic physics at play in the high-$T_c$'s 
is  a formidable challenge, 
both experimental and theoretical
investigations have shown that the magnetic
field-temperature ($B-T$) phase diagram of these materials
is also quite a bit more
complicated than for conventional superconductors \cite{BlatterRMP}. 
It is well known that the flux lines in a clean 
$s-$wave type-II superconductor form an Abrikosov triangular 
lattice for fields $B$ larger than the lower critical field, $B_{c_1}$.
However, for high-$T_c$'s, this mean field phase diagram is 
considerably affected by the 
combined effects of thermal fluctuations and various 
types of disorder, such as oxygen vacancies, 
columnar defects, or twin boundaries (see Fig. ~1a) \cite{BlatterRMP}.
\begin{figure}\label{Fig1}
\epsfxsize=7.5truecm
\centerline{\epsfbox{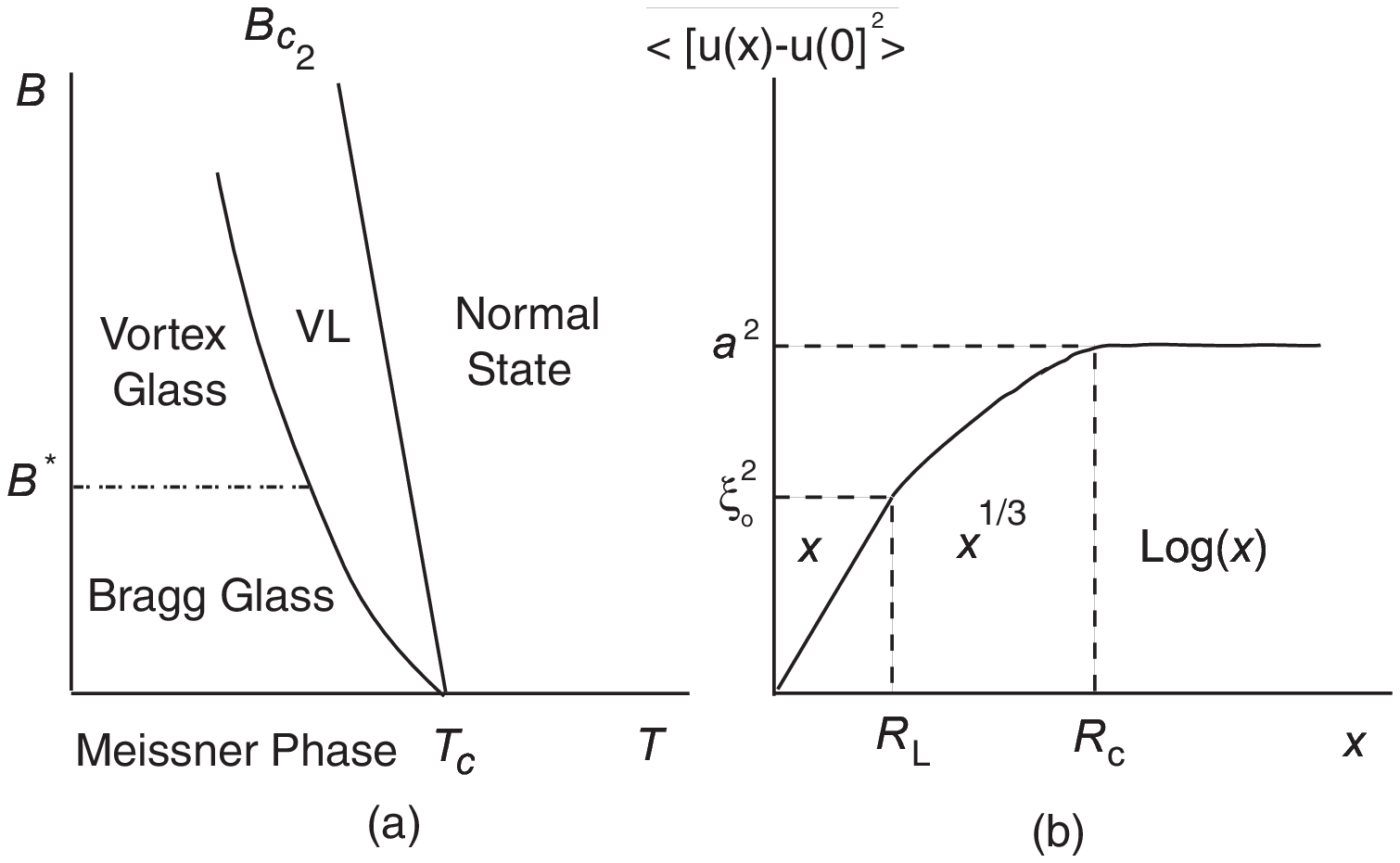}}
\end{figure}
\vspace{-5mm}
{\small{Fig.~1: (a) Schematic phase diagram of high 
temperature superconductors with point disorder. 
(b) The generic behavior of
the relative displacements correlation function.   
}}

Thermal fluctuations lead to melting of the 
vortex lattice and appearance of a vortex liquid (VL). 
In addition, extensive experimental, 
theoretical and numerical 
work strongly suggest that the Abrikosov vortex 
lattice in type-II superconductors is destroyed 
by the presence of disorder and gives place 
to various superconducting glass-like mixed states \cite{BlatterRMP}. 
If randomness is strong, the underlying translational order 
of the vortex lattice is completely destroyed
giving a vortex glass phase (VG).
However, for weak enough randomness, `some order'
is expected to survive. Indeed, it has been suggested that there may exist 
a stable, dislocation-free Bragg glass (BG) phase 
at low magnetic fields and temperatures in the 
presence of weak point impurities where the
structure factor $S({\bf k})$ has power law singularities at
 the Bragg reciprocal vectors ${\bf k} = {\bf G}$ \cite{GL1,GL2}.
For an applied field larger 
than a critical value $B^*$ (of order 2 to 20 
Teslas, depending on the specific sample considered)
\cite{Roulin}, the system undergoes a transition to a vortex glass
which consists of completely randomly frozen vortex 
lines with finite width peaks in the structure 
factor at  ${\bf k} \approx {\bf G}$ (see Fig.~1a) \cite{GL1,GL2,Ryu}.

The vortex state of high-$T_c$'s leads to a spatially varying magnetic field
${\bf B}({\bf r})$ inside the superconductor which may be investigating
using various techniques.
The muon
spin rotation ($\mu$SR) technique is an
almost ideal method for studying the
magnetic field distribution in the 
mixed state, and magnetic field and temperature 
dependence of $\lambda(B_0)$ and $\xi_0(B_0)$ \cite{muSR}. 
In a $\mu$SR experiment the inhomogeneous magnetic field
is sensitively monitored
by implanting muons into the sample, one at a time,
where the muon's spin precess with a Larmor frequency which 
is directly proportional to the local magnetic 
field ${\bf B}({\bf r})$. The muon-spin precession signal, which is 
obtained by detecting the muons positron-decay 
pattern, reflects the distribution of the Larmor
precession frequencies from 
all the muon stopping sites, hence forming an asymmetric
$\mu$SR lineshape \cite{muSR}. The width 
of this lineshape and, correspondingly, the width of the
internal field distribution, is
roughly proportional to $1/\lambda^2$.
A parametrization of the experimental lineshape based
on a theoretical 
lineshape produced by an effective London model that includes both
the cut-off effects arising from the finite-size of the vortex core and 
a phenomenological description of the effects of broadening caused
by static  and random distortion of the vortex lattice and/or nuclear
dipolar magnetic fields allows for a determination of effective 
$\lambda^{\rm eff} (B_0)$ and $\xi_0^{\rm eff} (B_0)$ 
\cite{muSR,Sonier94,Sonier99}. 

$\mu$SR experiments 
on YBa$_2$Cu$_3$O$_{6+\delta}$(YBCO) \cite{muSR,Sonier99} 
have reported a 
magnetic field $B_0$ dependence of the $\mu$SR 
lineshape. This has been so far attributed to a
microscopic and intrinsic magnetic field dependence 
of the London penetration length $\lambda$ and 
coherence length $\xi_0$ that could possibly be due 
to the underlying $d-$wave symmetry of the superconducting 
pairing state of YBCO \cite{Amin}. This brings up the
following question: Could the magnetic 
field dependence of the $\mu$SR lineshape and the
 field dependence ascribed to the extracted
effective $\lambda^{\rm eff}(B_0)$ 
and $\xi_0^{\rm eff}(B_0)$ be partially extrinsic and unrelated to 
the underlying microscopic $d-$wave physics, but rather
due to the fact that the lineshape ought to change from (somewhat)
asymmetric in the Bragg 
glass state to more or less a Gaussian shape 
in the vortex glass phase? 
Indeed, such field-driven evolution of the asymmetry (referred to
as skewness) of the $\mu$SR lineshape has been
reported for BSCCO \cite{Lee,Cubitt} and YBCO$_{6.60}$ \cite{SonierPRB}
high-$T_c$ materials. In this paper, we study the 
applied magnetic field $B_0$ dependence of the variance
of the field distribution, and consequently 
$\lambda^{\rm eff}(B_0)$, caused by weak point disorder. 
To this end, we use replica variational solutions
of the displacement correlations of the flux
lines in the presence of  weak
point disorder, 
and calculate the magnetic field variance, 
which is related to the structure factor $S({\bf k})$ of
the vortex system.  We note, however, that the
vortex-vortex displacement correlation functions of the
weakly disordered vortex lattice is 
anisotropic and may have nonuniversal exponents
\cite{Nattermann}. Consequently, we use an isotropic model to
simplify calculations and expose the basic underlying physics at stake.
Here, we only study the effect of disorder; the role of thermal 
fluctuation effects will be discussed elsewhere.

The influence of the randomness 
on the translational correlation function
has been discussed by many authors\cite{BlatterRMP}.
It was originally discussed by Larkin 
\cite{Larkin} using a model in which weak
random forces act independently on each
vortex. This model predicts an exponential
decay of the translation correlations 
on length scales larger than a disorder
dependent Larkin length, $R_L$. 
However, Bouchaud {\it et al.} pointed out that
at large scales the lattice starts
behaving collectively as an elastic manifold
in a random potential with many metastable
states and that the exponential decay does not
hold beyond the Larkin length $R_L$ \cite{BMY}. 
They used a variational replica field theory to study
the pinning of vortex lattice by impurities
they found a power law roughening of the 
lattice with stretched exponential decay
of the translational correlation function
for length scale beyond $R_L$ \cite{BMY}.
Later, Giamarchi and Le Doussal \cite{GL1}
showed that while disorder produce algebraic
growth of displacement correlations at short scales,
periodicity takes over at large scales and 
results in a growth of displacements that 
is at most logarithmic in $x$.  The results for the
displacement correlation function
are summarized in Fig.~1 (b). For length
scales smaller than $R_L$ the model is
equivalent to Larkin's model with correlations
$\propto x$ (in three dimension) and which corresponds to 
the replica symmetric part of the variational 
solution. For $R_L<x<R_c$, the system is in the random 
manifold (RM) regime \cite{BMY}. 
In this regime, replica symmetry is 
broken due the various metastable 
vortex configuration states and relative displacements correlate 
as $x^{2\nu}$, where $\nu=1/6$ in variational
approximation. For $x>R_c$, the periodicity 
of the lattice becomes important and one
enters the asymptotic Bragg glass logarithmic regime \cite{GL1,GL2}.

In the London limit, the
field of point vortices sitting at arbitrary
positions $r_m=(x_m,y_m,z_m)$ for $\lambda>>d$,
where $d$ is the distance between layers, 
varies slowly between layers, and is
given by \cite{Fingelmann,Others,BrandtPRL}:
\begin{eqnarray}\label{Br}
&&{\bf B}({\bf r})=\int \frac{d^3 k}{(2\pi)^3}
{\bf b}({\bf k}) \sum_m \exp\left[i{\bf k}.(
{\bf r}-{\bf r}_m)\right] \nonumber \\
&&{\bf b}({\bf k})=\frac{d\phi_0
(\hat z k_\perp^2-{\bf k_\perp} k_z)}{
k_\perp^2(1+\lambda^2 k^2)} \;\;\; ,
\end{eqnarray}
where ${\bf k}=({\bf k_\perp},k_z)$. 

Brandt \cite{BrandtPRL} had used Eq.~(\ref{Br}) 
to study the magnetic field variance in
layered superconductors. Brandt showed 
that perturbation of a lattice of 
rather stiff flux lines increase 
the field variance \cite{BrandtJLTP}.
However, the fluctuations of vortex line segments or 
vortex pancakes of highly flexible flux lines may 
decrease the $\mu$SR linewidth 
\cite{BrandtPRL,Fisher}. 

The general expression for the magnetic
field variance, $\delta^2\equiv\langle 
{\overline{[{\bf B}({\bf r})-[{\bf B}({\bf r})]]^2}}\rangle$, 
where $[...]$, $<...>$, and $\overline{...}$
denote space, thermal, and disorder average,
respectively, can be written as
\begin{equation}\label{delta1}
\delta^2=
\overline{\left\langle \int \frac{d^3 k}{(2\pi)^3}
\mid {\bf b}({\bf k})\mid^2 \frac{1}{V} 
\mid \sum_m \exp(-i{\bf k}.{\bf r}_m)\mid^2 \right \rangle}
\;\;\;,
\end{equation}
where $V$ is the volume of the sample. The 
lattice sum is the structure factor of the
vortex-point arrangement. One can rewrite Eq.~(\ref{delta1}) as
\begin{equation}\label{delta2}
\delta^2=\frac{\rho_0}{d}\int \frac{d^3 k}{(2\pi)^3}
\mid {\bf b}({\bf k})\mid^2 S({\bf k}),
\end{equation}
where $\rho_0=B_0/\phi_0$ is the density of vortices
and the structure factor $S({\bf k})$ is given by
\begin{equation}\label{SF}
S({\bf k})=\left(\frac{\rho_0}{d}\right)\sum_G \int d^3x
\hspace{0.1cm}
e^{i({\bf k_\perp}-{\bf G}).{\bf x}} C_G(x) \;\;\; .
\end{equation}
Here $G$ is reciprocal lattice vector and $C_G(x)$
is the translational correlation
function, which for simple isotropic case has the
following form \cite{GL1,GL2}
\begin{equation}\label{CGx}
C_G(x)=\exp\left\{\frac{-G^2}{2}\langle
{\overline{[u(x)-u(0)]^2}}\rangle\right\} \;\;\; ,
\end{equation}
where $u(x)$ is the displacement of the
flux line from its equilibrium position.

The above equation for $\delta^2$ has in general 
both perpendicular, $\delta_\perp$, and $z$,
$\delta_z$,  components. 
Before analyzing the effect of impurities
on the magnetic field variance, we first discuss
a  simple limiting case to estimate
the order of different components.
If the 
point vortices in each layer are assumed to be randomly 
positioned, then one simply has
\begin{eqnarray}\label{delta3}
\delta^2=\int \frac{d^3 k}{(2\pi)^3}
\frac{(B_0\phi_0 d)}{(1+\lambda^2 k^2)^2}\Bigg[1+
\frac{1}{(k_\perp^2+\lambda_J^{-2})}\Bigg],
\end{eqnarray}
The first term gives the fluctuations of $B_z$,
and the second term the fluctuations of ${\bf B}_\perp$.
A cutoff has been used due to the
factor $1/k_\perp^2$, for which the Josephson 
length $\lambda_J=d\lambda_c/\lambda$ was inserted 
by replacing $1/k_\perp^2$ by $1/(k_\perp^2+\lambda_J^{-2})$
\cite{BrandtPRL}. Integrating over ${\bf k}$ leads to
\begin{equation}\label{delta4}
\delta_\perp ^2=\frac{B_0 d\phi_0}{8\pi\lambda^3}\left\{1+
\frac{\cos^{-1}(\lambda_J/\lambda)}{\sqrt{(\lambda/\lambda_J)^2-1}}
\right\}.
\end{equation}
For $\lambda_J<<\lambda$ the second term which comes
from the random perpendicular fluctuations of ${\bf B}({\bf r})$ 
is small. Consequently, and also for sake of simplicity, the perpendicular
contribution shall be disregarded in the rest of the paper. 

Returning to the general expression Eq. (2), 
using Eqs. (3), (4) and (5), and considering only the $z$ component, and 
integrating first over all (six) angles and $k$ we find
\begin{equation}\label{delta2N}
\delta^2 \approx \delta_z^2=\frac{B_0^2}{2 \lambda^3} \sum_G
\frac{1}{G}\int_0^\infty x e^{-x/\lambda}C_G(x) 
\sin(G x) dx.
\end{equation}
The above equation can then be used to study
the variance of the magnetic field. 
The simple case of $C_G(x)=1$ corresponds
to the zero disorder case, in which integrating
over $x$ leads to the London model with
$\delta^2=B_0^2\sum_{G\neq 0}1/(1+\lambda^2 G^2)^2$.
For finite disorder, recalls that 
the relative displacement correlation function
has different behavior in the Larkin, RM, and 
logarithmic regimes (Fig. 1b). 
Therefore, we have to carry the integration of Eq.~(\ref{delta2N}) for 
these three $x-$dependent regime of $C_G(x)$.

Using expression (9) of Ref. \cite{GL2} for $R_c$,
\begin{equation}\label{Rc1}
R_c=\frac{2a_0^4c_{66}^{3/2}c_{44}^{1/2}}{\pi^3\rho_0^2 U_p^2 2\pi d \xi_0},
\end{equation}
where $a_0$ is the lattice spacing, $\xi_0$ is superconducting 
coherence length, $U_p$ is a typical pinning energy 
per unit length along $z$, and $c_{44}$ and $c_{66}$ are
elastic constants. Considering $c_{44}=c\varepsilon_0/(\gamma^2 a_0^2)$ and 
$c_{66}=\varepsilon_0/(4 a_0^2)$ with the vortex line tension
$\varepsilon_0=(\phi_0/4\pi\lambda)^2$, and $c=\ln(\gamma d/\xi_0)$,
$\gamma$ is the anisotropy ratio, one gets
\begin{equation}\label{Rc2}
R_c=\frac{2c^{1/2}}{\gamma d \xi_0^2}
\left(\frac{a_0}{2\pi}\right)^4
\left(\frac{\varepsilon_0}{U_p}\right)^2.
\end{equation}
The Larkin length is also given by $R_L\approx (\xi_0/a_0)^4 R_c$.
Taking parameters for YBCO
$\gamma=5$, $\lambda=1200A$, $d=10A$,
$\xi_0=10A$, and $U_p/\varepsilon_0=1/40$,
we see that $R_L\approx 5 \AA$ and 
$R_c\approx 10^7 \AA$ (for $B_0\sim 1T$). 
Since $R_L<<a_0$, $R_c>>a_0$, and $R_c >> \lambda$,
the effects of  Larkin and logarithmic 
regimes are very small,  the main 
contribution to the integral Eq. (8) comes from 
the RM regime as we have confirmed by explicit calculation.
Since the flux line displacement correlations
$\overline{<[u(x)-u(0)]^2>}\approx \xi_0^2(x/R_L)$ in the Larkin
regime and $\approx [\log(A x/R_c)+B]$ in the logarithmic regime ($A$ and $B$
are constants \cite{GL1}), $\delta$ can be calculated
exactly in those two regimes and 
the contributions of these two regimes are indeed negligible.

We note that using these numerical values,
the crossover field $H_{cross}=\pi c\phi_0/(\gamma^2 d^2)$
is of order of $10^2T$ and the Bragg glass to vortex glass
transition magnetic field
$H_M(T=0)=((\pi c_L)^4/(16\pi)^{1/3} \pi^2))(\epsilon_0/Up)^2
H_{c_2}^{3/2} H_{cross}^{1/3}$, where $c_L$ is the Lindeman
criterion, is $\sim 12T$ (for $H_{c_2}=100T$ and $c_L=0.12$)
\cite{GL1,GL2}.
Similar values but $U_p/\varepsilon_0=1/60$ leads to
$H_M(T=0)\approx 20T$, and if $\gamma=35$ and $U_p/\varepsilon_0=1/40$
one obtains $H_M(T=0)\approx 4T$ and $H_{cross}\approx 18T$ (we have
used these three different cases in Fig.~2).

Focusing now on the RM regime contribution to $\delta$, 
the displacement correlation function 
$\langle{\overline{[u(x)-u(0)]^2}}\rangle=
(a_0/\pi)^2(x/R_c)^{(2/\nu)}$,
where $\nu=1/6$ \cite{GL1,GL2,BMY}.
The magnetic field variance can then be calculated from Eq. (8):
\begin{equation}\label{delta2RM}
\delta_{RM}^2=
\sum_G\frac{B_0^2}{2 \lambda G}
\int_a^b y
{\rm e}^{[-\eta_G(\frac{\lambda}{R_c} y)^{1/3}-y]}
\sin(\lambda G y) dy,
\end{equation}
where $a=R_L/\lambda$, $b=R_c/\lambda$, 
$\eta_G=8G^2/(3K_0^2)$. The behavior of the magnetic field variance
(from numerical integration) is shown in Fig.~2.

One notes that the magnetic field
variance increases by increasing either magnetic  
field or the strength of the impurities. Since
$R_L$ is very small and $R_c$ is very large,
we can approximately replace the limit of integral
from $0$ to $\infty$, and do the integral analytically.
The results of analytical integration show the same
behavior for the magnetic field variance, in agreement
with numerical integration.

\vspace{-15mm}
\begin{figure}\label{Fig2}
\epsfxsize=7.5truecm
\centerline{\epsfbox{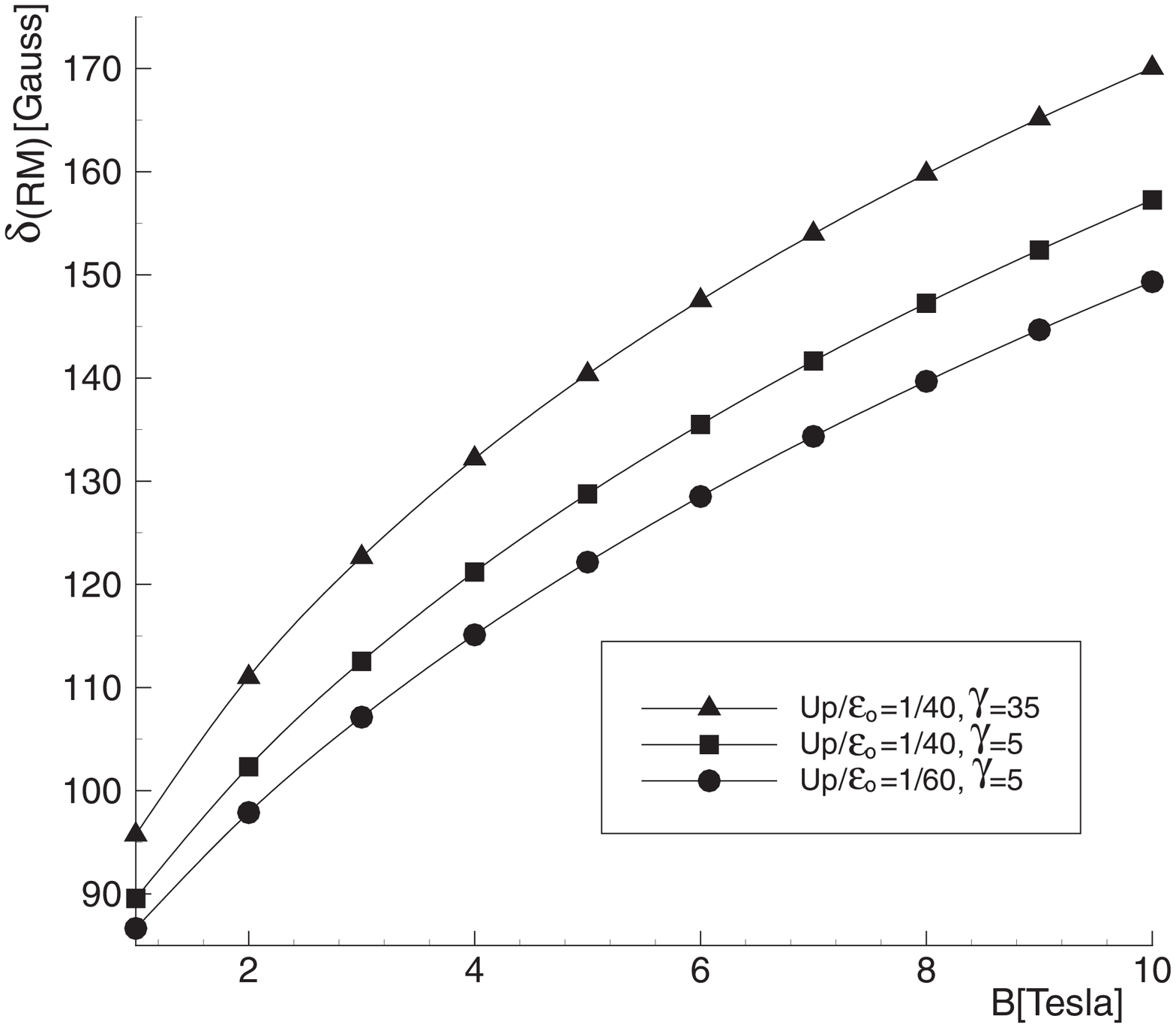}}
\end{figure}   
{\small{Fig.~2: Behavior of the variance $\delta$ of
the internal magnetic field as a function of the strength of 
the applied magnetic
field for different strength of the disorder.}}

We now briefly discuss our results in the context of
$\mu$SR measurements \cite{muSR,Sonier99}.
In Ref. \cite{Sonier99}, an {\it increase} of 
$\lambda^{\rm eff}(B_0)$ at
$T=4$ K between 25$\AA$ to 50$\AA$ per Tesla 
was reported.
This corresponds to an effective narrowing of 
$\mu$SR linewidth. Above, in Fig. 2, we found 
a {\it broadening}  of the 
internal magnetic field distribution 
by approximately 5$-$10 Gauss per Tesla of 
applied field $B_0$.
To compare the magnitude of our field induced linewidth effect 
to the experiment, we invoke the formula 
for $\delta$ that applies to that of a perfect
Abrikosov triangular vortex lattice in the
London limit 
$\delta_{\rm AVL} \sim  0.061 \Phi_0/\lambda^2$, where
$\Phi_0=2\times 10^{-15}$ Tm$^2$  \cite{muSR}.
We find that an increase of $\lambda^{\rm eff}$ 
of between
25$\AA$ to 50$\AA$ reported in Fig. 3
of Ref. \cite{Sonier99} corresponds to a decrease 
of the width of the
internal magnetic field distribution 
by about 4$-$8 Gauss per Tesla of applied field.
In other words, the effect we report is of 
the same magnitude
as that reported in $\mu$SR experiments \cite{Sonier99}
but ascribed therein to the intrinsic
nature of the field dependence of the microscopic $\lambda$.
The fact that we get a contribution of opposite 
trend (i.e. broadening) to that observed in experiments is irrelevant. 
Our results suggests
that the extrinsic modification of the internal 
field distribution as
the system evolves from the Bragg glass to the 
vortex glass is of the same order
as a contribution that may be of microscopic origin. 
Consequently,
a quantitative description of the $\mu$SR data that 
incorporates both
intrinsic and extrinsic effects is desirable in order 
to expose in a quantitative
manner any microscopic effects such as those discussed 
in Refs. \cite{Yip} and \cite{Amin}.
As mentioned above, the $\mu$SR  data are typically parametrized by
including a phenomenological Gaussian broadening 
parameter \cite{muSR,Sonier94,Sonier99}.
Possibly, the phenomenology exposed in our calculation 
above may already be handled to some extend by the usage of a 
phenomenogical Gaussian broadening parameter. However, 
 more theoretical, numerical and 
experimental work is needed to assess whether or not this is the case.
To conclude, we have shown that there can be 
significant magnetic field dependence    
for the magnetic field variance in the Bragg phase 
which is extrinsic and disorder driven
and has no microscopic $d-$wave origin. We hope that our 
preliminary study will motivate
further work.

We thank T. Giamarchi for useful discussions.
This work has been supported by NSERC of Canada, 
Research Corporation, and the Province of Ontario.

\end{multicols}
\end{document}